\documentclass{PoS}
\usepackage{amsmath}
\usepackage{cite}

\newcommand{\ON}[1]{\mathrm{O}( #1 )}
\newcommand{\SU}[1]{\mathrm{SU}( #1 )}
\newcommand{\SL}[1]{\mathrm{SL}( #1 )}
\newcommand{\GL}[1]{\mathrm{GL}( #1 )}
\newcommand{\SO}[1]{\mathrm{SO}( #1 )}

\newcommand{\USp}[1]{\mathrm{USp}(#1)}

\newcommand{\TA}{{\cal A}}
\newcommand{\TB}{{\cal B}}

\newcommand{\D}{\mathfrak{D}}

\newcommand{\tomega}{\tilde{\omega}}

\newcommand{\obf}[1]{\overline{\mathbf{#1}}}
\newcommand{\mbf}[1]{\mathbf{#1}}
\newcommand{\gL}{\mathcal{L}}
\newcommand{\hA}{\hat{A}}

\newcommand{\gM}{\mathcal{M}}
\newcommand{\gH}{\mathcal{H}}

\title{Half-maximal consistent truncations using exceptional field theory}

\ShortTitle{Half-maximal consistent truncations using EFT}

\author{\speaker{Emanuel Malek}\\
       Arnold Sommerfeld Center for Theoretical Physics, Ludwig-Maximilians-Universit\"at M\"unchen\\
       E-mail: \email{e.malek@lmu.de}}

\abstract{We show how to construct half-maximal consistent truncations of 10- and 11-dimensional supergravity to seven dimensions using exceptional field theory. This procedure gives rise to a seven-dimensional half-maximal gauged supergravity coupled to $n$ vector multiplets, with $n \neq 3$ in general. We also show how these techniques can be used to reduce exceptional field theory to heterotic double field theory.}

\FullConference{Corfu Summer Institute 2016 "School and Workshops on Elementary Particle Physics and Gravity"\\
         31 August - 23 September, 2016\\
         Corfu, Greece}

\begin{document}

\section{Introduction}
Exceptional field theory \cite{Berman:2010is,Hohm:2013vpa} is an extension of 10- and 11-dimensional supergravity, which treats the metric and $p$-form fields in a $d$-dimensional ``internal space'', on an equal footing, and has an extended set of coordinates. These features allow the exceptional field theory to make an $E_{d(d)}$ symmetry of the supergravity manifest.

This $E_{d(d)}$ symmetry is often confused with the U-duality group arising by compactifying 11-dimensional supergravity on a $T^d$. However, here the $E_{d(d)}$ symmetry arises before any truncation or compactification occurs, or equivalently all modes of the $d$-dimensional internal space are kept. Instead, the metric and $p$-form fields \emph{at each point} combine into $E_{d(d)}$ representations. We will see later how the global structure of the internal space determines the resulting ``duality group'', which to be precise we take in this case to be the global symmetry of the lower-dimensional gauged SUGRA that arises after truncating the theory.

In addition to repackaging the fields into $E_{d(d)}$ representations, the exceptional field theory extends the coordinates of the internal space to form a particular representation of $E_{d(d)}$. For the special case that the internal space is a torus, these coordinates have a nice interpretation as momentum and wrapping modes of membranes on $T^d$. For general backgrounds such an interpretation fails, but the extra coordinates allow us to treat 11-dimensional and 10-dimensional type II -- and indeed, as we will see, also heterotic -- supergravities at the same time. This allows us to easily see when a lower-dimensional theory can have different higher-dimensional origins, i.e. when there is a duality, and it is in this sense that one can think of exceptional field theory as making dualities manifest.\footnote{Note that with this interpretation in mind, we do not need to define the exceptional field theory structures over the enlarged space. Instead it is sufficient to define these structures only over a geometric manifold, corresponding to a subspace of the full enlarged space. If one wanted to consider non-geometric backgrounds, one would have to take care to define the exceptional structure over the full enlarged space. However, here will not need to do this as we will only consider geometric set-ups.}

Exceptional field theory \cite{Berman:2010is,Hohm:2013vpa}, as well as double field theory \cite{Hull:2009mi} and generalised geometry \cite{Gualtieri:2003dx,Coimbra:2011ky,Coimbra:2011nw} have proven very powerful in finding consistent truncations of 10- and 11-dimensional supergravities \cite{Aldazabal:2011nj,Geissbuhler:2011mx,Grana:2012rr,Dibitetto:2012rk,Berman:2012uy,Hohm:2014qga,Lee:2014mla,Lee:2015xga,Baguet:2015sma,Malek:2015hma,Baguet:2015iou}. This has allowed an uplift of various, previously ``orphaned'', gauged supergravities. However, so far this has focused on so-called generalised Scherk-Schwarz Ans\"atze which yield consistent truncations preserving all supersymmetries.

In \cite{Malek:2016bpu} and \cite{Malek:2016vsh} we considered exceptional field theory on backgrounds with non-trivial structure groups. The particular class of backgrounds considered admit half the possible number of spinors and thus have a generalised ${\cal N}=2$ structure. We showed how exceptional field theory naturally describes such backgrounds and developed the technology to find consistent truncations to seven-dimensional half-maximal gauged SUGRA. Using these tools we also showed how exceptional field theory can be reduced to the heterotic double field theory when the internal space has generalised $\SU{2}$-structure.

In this proceedings article we will present these results. In section \ref{s:Review} we will give a self-contained introduction to the $\SL{5}$ exceptional field theory. We will then show in section \ref{s:SU2} how the $\SL{5}$ exceptional field theory describes half-maximal backgrounds. Section \ref{s:ConsTruncation} is devoted to the study of half-maximal consistent truncations while section \ref{s:Het} details how to obtain the heterotic double field theory from exceptional field theory. We conclude in section \ref{s:Conclusions}.

\section{Review of $\SL{5}$ exceptional field theory}\label{s:Review}

We begin with a brief review of the key features of the $\SL{5}$ exceptional field theory \cite{Berman:2010is,Hohm:2013vpa,Musaev:2015ces} which are necessary for our purposes. We consider 11-dimensional supergravity but breaking $\GL{11} \longrightarrow \GL{7} \times \GL{4}$. We will see that having made $\SL{5}$ symmetry manifest we will automatically also have obtained the equivalent reformulation of type IIB supergravity. With respect to this splitting we write $x^{\hat{\mu}} = \left(x^\mu, x^i\right)$ with $\hat{\mu} = 1, \ldots, 11$, $\mu = 1, \ldots, 7$ and $i = 1, \ldots, 4$. Decomposing the metric and $p$-form fields at each point we have
\begin{equation}
 \begin{split}
  g_{\hat{\mu}\hat{\nu}} &\longrightarrow g_{\mu\nu}, \,\, g_{\mu i}, \,\, g_{ij} \,, \\
  C_{\hat{\mu}\hat{\nu}\hat{\rho}} &\longrightarrow C_{\mu\nu\rho}, \,\, C_{\mu\nu i}, \,\, C_{\mu ij}, \, C_{ijk} \,, \\
  C_{\hat{\mu}_1\ldots \hat{\mu}_6} &\longrightarrow C_{\mu_1\ldots \mu_6} ,\,\, C_{\mu_1\ldots \mu_5 i} ,\,\, C_{\mu_1\ldots \mu_4 i j} ,\,\, C_{\mu_1 \mu_2 \mu_3 ijk} ,\,\, C_{\mu_1 \mu_2 ijkl} \,.
 \end{split}
\end{equation}

The 14 fully internal components $g_{ij}$, $C_{ijk}$ parameterise the coset space $\frac{\SL{5}}{\SO{5}}$ and can be combined into the so-called generalised metric
\begin{equation}
 \gM_{ab} \in \frac{\SL{5}}{\SO{5}} \,,
\end{equation}
where $a, b = 1, \ldots, 5$ are fundamental $\SL{5}$ indices. Similarly, the 10 components which transform as a vector under $\GL{7}$, i.e. $g_{\mu i}$, $C_{\mu ij}$ can be combined into a $\GL{7}$-covector valued in the $\mathbf{10}$ of $\SL{5}$
\begin{equation}
 \TA_\mu{}^{ab} = \TA_\mu{}^{[ab]} \,.
\end{equation}
One can continue and combine the five $C_{\mu\nu i}$ and $C_{\mu\nu ijkl}$ into an object $\TB_{\mu\nu,a}$, and so on.

Now that the metric and $p$-form fields have been combined into $\SL{5}$ representations, we also extend the four-dimensional internal space to be ten-dimensional, with coordinates $Y^{ab} = Y^{[ab]}$. This will allow us to write fully $\SL{5}$ covariant expressions, even though as we will see not all of these coordinates are physical. For example, diffeomorphisms and $p$-form gauge transformations combine into a local $\SL{5}$ transformation. These symmetries are generated by the generalised Lie derivative \cite{Berman:2011cg,Coimbra:2012af,Berman:2012vc} which acts on a $\SL{5}$ vector $V^a$ with weight $w$ as
\begin{equation}
 \gL_\Lambda V^a = \Lambda^{bc} \partial_{bc} V^a - V^b \partial_{bc} \Lambda^{ac} + \left( \frac{w}{2} + \frac15 \right) V^a \partial_{bc} \Lambda^{bc} \,, \label{eq:gL}
\end{equation}
where $\Lambda^{ab} = \Lambda^{[ab]}$ is a so-called ``generalised vector field'' and has weight $\frac15$. The derivatives $\partial_{ab}$ are with respect to the extended coordinates $Y^{ab}$.

These generalised diffeomorphism symmetries must close into an algebra and this requires the following condition on all pairs of fields which we denote abstractly as $f$ and $g$
\begin{equation}
 \partial_{[ab} f \partial_{cd]} g = \partial_{[ab} \partial_{cd]} f = 0 \,.
\end{equation}
This is known as the ``section condition'' \cite{Berman:2011cg} and restricts all fields to depend only on a subset of coordinates. Up to $\SL{5}$ transformations there are two solutions:
\begin{equation}
 \begin{split}
  \mathrm{(i)} &\quad \partial_{ij} = 0 \,, \quad \partial_{i5} \neq 0 \,, \qquad \qquad ~\,\, \textrm{for} \quad i = 1, \ldots, 4 \,, \\
  \mathrm{(ii)} &\quad \partial_{\alpha 4} = \partial_{\alpha 5} = 0 \,, \quad \partial_{\alpha \beta} \neq 0 \,, \quad \textrm{for} \quad \alpha = 1, 2, 3 \,.
 \end{split}
\end{equation}
The first clearly breaks $\SL{5} \longrightarrow \SL{4}$ while the second breaks $\SL{5} \longrightarrow \SL{3} \times \SL{2}$. It should thus not come as a surprise that the first corresponds to 11-dimensional supergravity while the second to type IIB supergravity \cite{Blair:2013gqa,Hohm:2013vpa}. One way to see this is to act with the generalised Lie derivative on the generalised metric $\gM_{ab}$. With the corresponding solution to the section condition the generalised Lie derivative reduces to diffeomorphisms and the correct $p$-form gauge transformation for these two theories.

Equipped with the generalised Lie derivative \eqref{eq:gL} one can introduce a generalised connection $\nabla_{ab}$ such that
\begin{equation}
 \nabla_{ab} V^c \equiv \partial_{ab} V^c + \Gamma_{ab,c}{}^d V^d \,,
\end{equation}
for a generalised tensor $V^c$ is also generalised tensor. This implies a certain transformation law for the components $\Gamma_{ab,c}{}^d$, which means that they are not tensorial themselves. However, as usual, there are certain combinations of the connection which are by themselves tensors. These are the generalised torsion of the generalised connection, and are given by
\begin{equation}
 \gL_{\Lambda}^{\nabla} V^a - \gL \Lambda V^a = \frac12 \tau_{bc,d}{}^a \Lambda^{bc} V^d \,, \label{eq:Torsion}
\end{equation} 
where $\Lambda$ is any generalised vector and $\gL_{\Lambda}^\nabla$ denotes the generalised Lie derivative with all derivatives replaced by the connection $\nabla$. The definition \eqref{eq:Torsion} makes it manifest that the torsion is a tensor. Later on we will make use of the fact that the torsion takes values in the representations
\begin{equation}
 W \equiv \mbf{10} \oplus \mbf{15} \oplus \mbf{40} \subset \mbf{10} \otimes \mbf{24} = \mbf{10} \oplus \mbf{15} \oplus \mbf{40} \oplus \mbf{175} \,.
\end{equation}

We conclude this section by noting that it is possible to rewrite the 11-dimensional and type IIB supergravity actions in terms of $\gM_{ab}$, $\TA_\mu{}^{ab}$ and the other fields. This action is uniquely fixed by the requirement that it is invariant under the generalised Lie derivative, up to the section condition, as well as the external seven-dimensional diffeomorphisms, which we have not discussed here as they are not important for what follows. We will call the part of the action that has only ``internal'' derivatives with respect to the 10 $Y^{ab}$ as the potential, $V$. It is given by
\begin{equation}
 V = - \frac14 {\cal R} - \frac18 \gM^{ab} \gM^{cd} \nabla_{ab} g_{\mu\nu} \nabla_{cd} g^{\mu\nu} \,, \label{eq:EFTV}
\end{equation}
where ${\cal R}$ depends only on the generalised metric $\gM_{ab}$ and
\begin{equation}
 \nabla_{ab} g_{\mu\nu} = |e|^{2/7} \partial_{ab} \left( g_{\mu\nu} |e|^{-2/7} \right) \,.
\end{equation}
As we will show later upon performing a truncation this part correctly reproduces the scalar potential of seven-dimensional half-maximal gauged SUGRA, as well the internal part of the heterotic DFT.

\section{$\SU{2}$-structures and half-maximal supersymmetry in exceptional field theory} \label{s:SU2}
Let us now turn our attention to internal spaces which admit nowhere vanishing spinors corresponding to 16 supercharges. Because we wish to include fluxes these spinors transform in the fundamental of $\USp{4}$, not $\SO{4}$ as one would expect from the 11-dimensional perspective. A $\SU{2}_R \subset \USp{4}$ doublet of spinors carries the appropriate 16 supercharges and the existence of these spinors implies that the generalised structure group is reduced to $\SU{2}_S \subset \USp{4}$ which stabilises this $\SU{2}_R$ doublet \cite{Coimbra:2014uxa}. Here the subscripts $R$ and $S$ are used to distinguish the $\SU{2}$ R-symmetry from the structure group. We will call these internal spaces generalised $\SU{2}$-manifolds.

\subsection{Reformulating the exceptional field theory}
We now seek a bosonic description of generalised $\SU{2}$-manifolds. One can show \cite{Malek:2016bpu} that one can form the following generalised tensors as bilinears of the well-defined spinors:
\begin{equation}
 \kappa\,, \qquad A^a\,, \qquad \hat{A}_a\,, \qquad B_u{}^{ab}\,, \,,
\end{equation}
where $a = 1, \ldots, 5$ is a $\SL{5}$ index and $u = 1, \ldots, 3$ labels the triplet of $\SU{2}_R$. We will throughout raise and lower the $\SU{2}_R$ triplet indices $u$ by $\delta_{uv}$. Using Fierz identities one can show that these satisfy
\begin{equation}
 A^a \hat{A}_a = \kappa^5 \,, \qquad B_u{}^{ab} \hat{A}_b = 0 \,, \qquad \frac14 B_{u}{}^{ab} B_{v}{}^{cd} \epsilon_{abcde} = \hat{A}_e \,, \label{eq:Compatibility}
\end{equation}
where $\epsilon_{abcde} = \pm 1$. Here $B_u{}^{ab}$ is a generalised vector field, thus has weight $\frac15$ under the generalised Lie derivative \eqref{eq:gL}, while $\hat{A}$ has weight $\frac25$ and $A^a$ has weight $\frac35$ under the generalised Lie derivative. These objects should be thought of as the generalisation of complex and K\"ahler structures on K3. Readers familiar with exceptional field theory will recognise these objects from the tensor hierarchy of exceptional field theory \cite{Hohm:2015xna,Wang:2015hca}. This is not a coincidence since sections of the appropriate vector bundles behave in many ways as generalisations of differential forms \cite{Cederwall:2013naa}. Indeed, this observation allows one to generalise this construction to other dimensions \cite{Malek:2017njj}.

We will now show that the set of nowhere vanishing $\kappa$, $A^a$, $\hA_a$ and $B_u{}^{ab}$ define a generalised $\SU{2}$-structure. We will do so by showing that they are stabilised by a $\SU{2}_S \subset \SL{5} \times \mathbb{R}^+$ subgroup. We begin by noting that the scalar density $\kappa$ breaks the $\SL{5} \times \mathbb{R}^+$ structure group of the exceptional field theory to $\SL{5}$.

It is easy to show that $A^a$ and $\hat{A}_a$ subject to \eqref{eq:Compatibility} are stabilised by a $\SL{4} \subset \SL{5}$ subgroup and thus nowhere vanishing $A^a$ and $\hat{A}_a$ define a $\SL{4} \simeq \SO{3,3}$ structure. Upon performing a consistent truncation, $A^a$ and $\hat{A}$ will lead to the dilaton scalar field of the seven-dimensional gauged SUGRA, and thus we will also call a set of globally well-defined nowhere-vanishing $A^a$ and $\hat{A}_a$ a ``dilaton structure''.

Furthermore, a set of three globally well-defined nowhere-vanishing $B_u{}^{ab}$ subject to \eqref{eq:Compatibility}, further reduce the generalised structure group to $\SU{2}_S \subset \SL{4} \subset \SL{5}$. Because $\SU{2}_S \subset \USp{4}$ and the generalised metric is a generalised $\USp{4}$-structure, the objects $A^a$, $\hat{A}_a$ and $B_u{}^{ab}$ together implicitly define a generalised metric. However, there is in general no explicit expression for the generalised metric in terms of $A^a$, $\hat{A}_a$ and $B_u{}^{ab}$. Nonetheless, because they carry the same degrees of freedom, it is possible to express the exceptional field theory action in terms of $A^a$, $\hat{A}_a$ and $B_u{}^{ab}$ as we will demonstrate in the next section.

It is also useful to further define the following objects using the generalised $\SU{2}$-structure
\begin{equation}
 V_{u,ab} = \frac14 \epsilon_{abcde} B_u{}^{cd} A^e \,, \qquad K_{uv}{}^a{}_b = \kappa^{-5} B_{[u}{}^{ac} V_{v]bc} \,,
\end{equation}
where $V_{u,ab}$ is in the $\obf{10}$ of $\SL{5}$ and has weight $\frac45$, while $K_{uv}{}^a{}_b$ is in the adjoint of $\SL{5}$. It is easy to show that it satisfies the $\SU{2}_R$ algebra
\begin{equation}
 \left[ K_{uv}, K_{wx} \right]^a{}_b = \delta_{u[w} K_{x]v}{}^a{}_b - \delta_{v[w} K_{x]u}{}^a{}_b \,,
\end{equation}
and acts on $B_u{}^{ab}$ as
\begin{equation}
 2 K_{uv}{}^{[a}{}_c B_{w}{}^{b]c} = \delta_{w[u} B_{v]}{}^{ab} \,.
\end{equation}
From this one can see that $K_{uv}{}^a{}_b$ generates the $\SU{2}_R \subset \SL{5}$ subgroup. Furthermore, in the case of 11-dimensional supergravity, it becomes the hypercomplex structure on the ``internal'' four-manifold.

\subsection{Intrinsic torsion}
The first step in expressing the exceptional field theory action in terms of $A^a$, $\hat{A}_a$ and $B_u{}^{ab}$ is to introduce a generalised connection. The natural choice here is given by a generalised $\SU{2}$ connection which means that it is compatible with the generalised $\SU{2}$-structure, i.e.
\begin{equation}
 \nabla_{ab} A^c = \nabla_{ab} \hat{A}_c = \nabla_{ab} B_u{}^{cd} = \nabla_{ab} \kappa = 0 \,. \label{eq:NablaCompatibility}
\end{equation}
Note that we are not imposing a torsion constraint on this connection and so it will certainly not be unique. However, we will not require the connection explicitly. We will only need to make use of the relations \eqref{eq:NablaCompatibility}.

To write the action we want to use generalised tensors which are given by one derivative of the $\SU{2}$-structure. We use the \emph{intrinsic torsion} of a $\SU{2}$ connection. This is the part of the torsion which is independent of the choice of $\SU{2}$ connection. This implies that it can be written without a $\SU{2}$ connection appearing explicitly. Furthermore, it should only involve the $\SU{2}$-structure. One can easily show that the intrinsic torsion consists of the following representations
\begin{equation}
 W_{int} = 2 \cdot \left(\mbf{1},\mbf{1}\right) \oplus \left(\mbf{3},\mbf{1}\right) \oplus 2 \cdot \left(\mbf{1},\mbf{3}\right) \oplus \left(\mbf{3},\mbf{3}\right) \oplus 3 \cdot \left(\mbf{2},\mbf{2}\right) \oplus \left(\mbf{2},\mbf{4}\right) \,. \label{eq:IntrinsicTorsionReps}
\end{equation}

We can give explicit expressions for the intrinsic torsion by making use of the generalised Lie derivative. We begin by considering
\begin{equation}
 \gL_{B_u} B_v{}^{ab} = \gL_{B_u}^\nabla B_v{}^{ab} + B_u{}^{cd} \tau_{cd,e}{}^{[a} B_v{}^{b]e} = B_u{}^{cd} \tau_{cd,e}{}^{[a} B_v{}^{b]e} \,,
\end{equation}
where we have made use of the fact that $\nabla$ is an $\SU{2}$-connection to show that this is an element of the $\SU{2}$ torsion. It is clear that it is independent of the choice of $\SU{2}$-connection because the left-hand side does not make use of a $\SU{2}$-connection. Thus it is intrinsic. Similarly, one can show that
\begin{equation}
 \gL_{B_u} A^a \,, \qquad \gL_{B_u} \kappa^5 \,, \qquad \partial_{ba} A^b \,,
\end{equation}
are also intrinsic. Keeping track of the representations that appear and with the help of some algebra one can write \cite{Malek:2017njj}
\begin{equation}
 \begin{split}
  \gL_{B^u} B_u{}^{ab} &= \kappa^2 T^{ab} + \kappa T^u B_u{}^{ab} + \kappa^{-1} T^{[a} A^{b]} \,, \\
  \gL_{B_{[u}} B_{v]}{}^{ab} &= \kappa^2 R_{uv}{}^{ab} + \kappa R_{uvw} B^{w,ab} - 2 \kappa^{-1} T^c K_{uv}{}^{[a}{}_c A^{b]} + \frac23 \kappa T_{[u} B_{v]}{}^{ab} \,, \\
  \gL_{B_u} A^a &= \kappa^3 B_u{}^{ab} S_b + \kappa \left( U_u - \frac23 T_u \right) A^a + \kappa^4 S_u{}^a \,, \\
  \gL_{B_u} \kappa^5 &= \kappa^6 U_u \,, \\
  \partial_{ba} A^b &= \kappa^3 P_a + \kappa \hA_a P \,.
 \end{split}
\end{equation}

The generalised tensors $T^{ab}$, $T^u$, $T^a$, $R_{uv}{}^{ab}$, $R_{uvw}$, $S_a$,  $S_u{}^a$, $U_u$, $P_a$ and $P$, which for later convenience we have defined to have weight $-\frac15$, correspond to the irreducible representations of the intrinsic torsion. They are explicitly given by
\begin{equation}
 \begin{split}
  T^a &= 2 \kappa^{-4} \hA_b \gL_{B^u} B_{u}{}^{ab} \,, \qquad T_u = \kappa^{-6} V_u{}_{ab} \gL_{B^v} B_v{}^{ab} \,, \\
  T^{ab} &= \kappa^{-2} \gL_{B^u} B_u{}^{ab} - \kappa^{-1} B_u{}^{ab} T^u - \kappa^{-3} T^{[a} A^{b]} \,, \\
  R_{uvw} &= \kappa^{-6} V_{[w|ab|} \gL_{B_u} B_{v]}{}^{ab} \,, \\
  R_{uv}{}^{ab} &= \kappa^{-2} \gL_{B_{[u}} B_{v]}{}^{ab} - \kappa^{-1} R_{uvw} B^{w,ab} + 2 \kappa^{-1} T^c K_{uv}{}^{[a}{}_c A^{b]} + \frac23 \kappa T_{[u} B_{v]}{}^{ab} \,, \\
  S_a &= -\frac43 \kappa^{-3} V^u{}_{ab} \gL_{B_u} A^b \,, \qquad S_u{}^a = \kappa^{-4} \gL_{B_u} A^a - \kappa^{-1} B_u{}^{ab} S_b - \kappa^{-3} \left(U_u - \frac23 T_u \right) A^a \,, \\
  U_u &= \kappa^{-6} \gL_{B_u} \kappa^5 \,, \qquad P = \kappa^{-6} A^a \partial_{ba} A^b \,, \qquad P_a = \kappa^{-3} \partial_{ba} A^b - \kappa^{-2} \hA_a P \,,
 \end{split}
\end{equation}
and satisfy
\begin{equation}
 \begin{split}
  T^a \hA_a &= 0 \,, \qquad \hA_a T^{ab} = V^u{}_{ab} T^{ab} = 0 \,, \qquad R_{uv}{}^{ab} \hA_a = R_{uv}{}^{ab} V^w{}_{ab} \,, \\
  S_a A^a &= 0 \,, \qquad S_u{}^a \hA_a = S_u{}^a V^u{}_{ab} = 0 \,, \qquad P_a A^a = 0 \,.
 \end{split}
\end{equation}
This implies that they correspond to the following irreducible representations of $\SU{2}_S \times \SU{2}_R$
\begin{equation}
 \begin{split}
  T_u &\in \left(\mbf{1},\mbf{3}\right) \,, \qquad T^{ab} \in \left(\mbf{3},\mbf{1}\right) \,, \qquad T^a \in \left(\mbf{2},\mbf{2}\right) \,, \\
  R_{uv}{}^{ab} &\in \left(\mbf{3},\mbf{3}\right) \,, \qquad R_{uvw} \in \left(\mbf{1},\mbf{1}\right) \,, \qquad U_u \in \left(\mbf{1},\mbf{3}\right) \,, \\
  S_a &\in \left(\mbf{2},\mbf{2}\right) \,, \qquad S_u{}^a \in \left(\mbf{2},\mbf{4}\right) \,, \qquad P_a \in \left(\mbf{2},\mbf{2}\right) \,, \qquad  P \in \left(\mbf{1},\mbf{1}\right) \,.
 \end{split}
\end{equation}
These are the exactly the representations of the intrinsic torsion \eqref{eq:IntrinsicTorsionReps}.

One can show explicitly that any other generalised tensor involving one derivative and constructed from the $\SU{2}$-structure is given by a linear combination of the intrinsic torsion above. For example, one can show that
\begin{equation}
 \begin{split}
  \gL_{B_{(u}} B_{v)}{}^{ab} &= \frac13 \delta_{uv} \gL_{B^w} B_w{}^{ab} \,,
 \end{split}
\end{equation}
and thus the symmetric part of $\gL_{B_u} B_v{}^{ab}$ is fully determined by its trace and thus by the intrinsic torsion $T^{ab}$, $T_u$ and $T^a$.

\subsection{Reformulating the action}
One can reformulate the exceptional field theory action in terms of the $\SU{2}$-structure. For example, the kinetic terms of the generalised metric, are given by \cite{Hohm:2013vpa}
\begin{equation}
 L_{kin} = \frac14 g^{\mu\nu} \D_\mu \gM^{ab} \D_\nu \gM_{ab} \,, 
\end{equation}
where $\D_\mu = \partial_\mu - \gL_{A_\mu}$ is the $\SL{5}$-covariant derivative with respect to the external seven dimensions. This term can be written in terms of the $\SU{2}$-structure as
\begin{equation}
 L_{kin} = \frac14 \kappa^{-5} g^{\mu\nu} \D_\mu B_u{}^{ab} \D_\nu B^{u,cd} \epsilon_{abcde} A^e + \kappa^{-10} g^{\mu\nu} \hA_a \D_\mu A^a \D_\nu \kappa^5 \,. \label{eq:ScalarKinTerms}
\end{equation}
The kinetic terms for the field strengths can similarly be rewritten by replacing the generalised with the $\SU{2}$-structure.

The potential term in the action \eqref{eq:EFTV} can be rewritten in terms of the $\SU{2}$-structure as
\begin{equation}
 V = - \frac14 {\cal R} + \frac12 B_{u}{}^{ab} B^{u,cd} \nabla_{ab} g_{\mu\nu} \nabla_{cd} g^{\mu\nu} \,. \label{eq:Potential}
\end{equation}
By writing the $\SU{2}$-structure in terms of spinor bilinears, as detailed in \cite{Malek:2016bpu} one can show that ${\cal R}$ is given by
\begin{equation}
 \begin{split}
  \cal{R} &= -2 U_u U^u + \frac43 U_u T^u + \frac{2}{3} T^u T_u + \frac13 R_{uvw} R^{uvw} + \frac14 \kappa^{-3} R_{uv}{}^{ab} R^{uv,cd} \epsilon_{abcde} A^e \\
  & \quad + \frac{1}{18} T^{ab} T^{cd} \epsilon_{abcde} A^e - 4 \kappa^{-2} \gL_{B_u} \left( U^u \kappa \right) - \frac12 P^2 + \frac{\sqrt{2}}{3} P\, \epsilon_{uvw} R^{uvw} + \ldots \,, \label{eq:RicciScalar}
 \end{split}
\end{equation}
where $\ldots$ refers to terms consisting only of the doublets of the intrinsic torsion, which we will not need in the following.

\section{Half-maximal consistent truncations}\label{s:ConsTruncation}
So far we have discussed how to describe a general 10- or 11-dimensional supergravity background in exceptional field theory which admits half the full number of spinors. We now want to describe a truncation of the theory on such a background in order to obtain a half-maximal gauged SUGRA, and discuss the requirements for consistency. Here, a consistent truncation is one where any solution of the lower-dimensional gauged SUGRA is also a solution of the original higher-dimensional SUGRA.

A key feature in obtaining consistency will be to remove any doublets of $\SU{2}_S$ from the truncation. Any such mode would correspond to additional spinors on the background, i.e. the background would admit more than half-supersymmetry. In terms of supergravity fields removing these modes is equivalent to projecting out the massive gravitino multiplets associated to the broken supersymmetries.

\subsection{Truncation Ansatz}
We make an Ansatz for our truncation by expanding the $\SU{2}$-structure in terms of a set of objects which depend only on the internal coordinates $Y^{ab}$, and are related to the modes that we keep in the truncation. The coefficients of these objects are scalar functions of the external seven coordinates $x^\mu$ and become the scalars of the half-maximal gauged SUGRA. In order to have a half-maximal SUGRA we expand the $\SU{2}$-structure in such a way that there are no doublets of $\SU{2}_S$. To be explicit we take
\begin{equation}
 \begin{split}
  \langle A^a\rangle(x,Y) &= \rho^3(Y) n^a(Y) e^{-2d(x)} \,, \qquad \langle\hat{A}_a\rangle(x,Y) = \rho^2(Y) \hat{n}_a(Y) \,, \\
  \langle B_u{}^{ab}\rangle(x,Y) &= \rho(Y) \omega_M{}^{ab}(Y) b_u{}^M(x) \,, \qquad \langle\kappa\rangle(x,Y) = \rho(Y) e^{-2d(x)/5} \,, \label{eq:TruncationAnsatz}
 \end{split}
\end{equation}
and we make a warped Ansatz for the external seven-dimensional metric
\begin{equation}
 \langle g_{\mu\nu}\rangle(x,Y) = \rho^{2}(Y) \hat{g}_{\mu\nu}(x) e^{-4d(x)/5} \,.
\end{equation}
Here the angled brackets $\langle \rangle$ denote the truncated objects, $\rho(Y)$ is a scalar density and $\omega_M{}^{ab}(Y)$, $n^a(Y)$ and $\hat{n}_a(Y)$ satisfy
\begin{equation}
 \begin{split}
  n^a \hat{n}_a &= 1 \,, \qquad \omega_M{}^{ab} \hat{n}_a = 0 \,, \qquad \frac14 \omega_{M}{}^{ab} \omega_{N}{}^{cd} \epsilon_{abcde} = \eta_{MN} \hat{n}_e \,, \label{eq:ModeCompatibility}
 \end{split}
\end{equation}
and $M = 1, \ldots, n+3$ with $n$ in principle arbitrary and $\eta_{MN}$ has signature $\left(3, n\right)$. These objects are the half-maximal analogue of twist matrices in the generalised Scherk-Schwarz procedure \cite{Aldazabal:2011nj,Geissbuhler:2011mx,Grana:2012rr,Dibitetto:2012rk,Berman:2012uy,Hohm:2014qga,Lee:2014mla,Berman:2013uda,Blair:2014zba}. As we will see $n$ corresponds to the number of vector multiplets of the half-maximal gauged SUGRA that we obtain. Throughout this paper we will use $\eta_{MN}$ to raise and lower the $M, N$ indices.

The functions $d(x)$, $b_u{}^M(x)$ are the scalars of the lower-dimensional gauged SUGRA. Due to the compatibility condition \eqref{eq:Compatibility}, $b_u{}^M$ must satisfy
\begin{equation}
 b_u{}^M b_v{}^N \eta_{MN} = \delta_{uv} \,.
\end{equation}
This imposes six constraints on the $3n+9$ scalars $b_u{}^M$. Additionally, we will identify any scalars related by the action of $\SU{2}_R$. This removes another three degrees of freedom. The remaining $3n+1$ scalars $b_u{}^M$ and $d$ parameterise the coset space
\begin{equation}
 {\cal M}_{scalar} = \frac{\ON{3,n}}{\ON{3}\times\ON{n}} \times \mathbb{R}^+ \,.
\end{equation}
The coset structure can be made more explicit by writing
\begin{equation}
 b_{u,M} b^u{}_N = \frac12 \left( \eta_{MN} - \gH_{MN} \right) \,,
\end{equation}
where $\gH_{MN}$ satisfies
\begin{equation}
 \gH_{MP} \gH_{NQ} \eta^{PQ} = \eta_{MN} \,.
\end{equation}

It is useful to also define
\begin{equation} 
 \omega_{M,ab} = \frac14 \epsilon_{abcde} \omega_M{}^{cd} n^e \,, \qquad  \tomega_M{}^{ab} = \rho \omega_M{}^{ab} \,,
\end{equation}
where $\tomega_M{}^{ab}$ are generalised vectors. The truncation Ansatz for the gauge fields $\TA_\mu{}^{ab}$, $\TB_{\mu\nu,a}$, etc. makes use of the same modes $\rho$, $\omega_M{}^{ab}$, $n^a$ and $\hat{n}_a$ as determined by their $\SL{5}$ structure.

\subsection{Consistency conditions and embedding tensor}
Our discussion so far has been purely algebraic. We now need to impose a set of differential constraints in order to obtain a consistent truncation, and we do this via the intrinsic torsion. Firstly, we must make sure that derivatives of the $\SU{2}$-structure do not source doublets of $\SU{2}_S$ since we have removed these from the truncation Ansatz. This means that the doublets of $\SU{2}_S$ in the intrinsic torsion must vanish.

Secondly, we need to ensure that the finite set of modes that we do keep, i.e. $\rho$, $\omega_M{}^{ab}$, $\hat{n}_a$ and $n^a$ do not source other modes. Thus the intrinsic torsion should close into this set of modes. Together these two requirements imply that we can write
\begin{equation}
 \begin{split}
  \gL_{\tomega_{M}} \tomega_{N}{}^{ab} &= f_{MNP} \tomega^{P,ab} +  \eta_{MN} f^P \tomega_P{}^{ab} + 2 f_{[M} \tomega_{N]}{}^{ab} \,, \qquad \gL_{\tomega_M} \rho^5 = \rho^5 \xi_M \,, \\
  \gL_{\tomega_M} \left(\rho^3 n^a \right) &= \rho^3 n^a \left(\xi_M - 2  f_M \right) \,, \qquad \partial_{ba} \left(\rho^3 n^b \right) = \rho^2 \hat{n}_a \theta \,. \label{eq:DiffConstraints}
 \end{split}
\end{equation}
By the structure of the generalised Lie derivative $f_{MNP} = f_{[MNP]}$ is totally antisymmetric and the right-hand side is the most general allowed subject to the conditions discussed above. The objects $f_{MNP}$, $f_M$, $\xi_M$ and $\theta$ are exactly the right representations to form the embedding tensor of half-maximal gauged SUGRA \cite{Dibitetto:2015bia,Bergshoeff:2007vb} coupled to $n$ vector mutliplets, that is they satisfies the linear constraint gauged SUGRA. Indeed, we will see that this is the correct interpretation and thus the embedding tensor can be thought of as the intrinsic torsion of the $\SU{2}$-structure background.

The final consistency condition we need to impose that $f_{MNP}$, $f_M$, $\xi_M$ and $\theta$ are constant. This final condition is completely analogous to the case of maximally supersymmetric consistent truncation. It is important to highlight that the construction presented here naturally leads to the full embedding tensor including the deformation parameter $\theta$.

\subsection{Truncated intrinsic torsion and scalar potential}
Before evaluating the scalar potential with the truncation Ansatz let us first compute the intrinsic torsion. With the truncation Ansatz \eqref{eq:TruncationAnsatz} and the differential constraints \eqref{eq:DiffConstraints} one finds
\begin{equation}
 \begin{split}
  \langle T_u \rangle &= 3 \rho^{-1} e^{2d/5} b_u{}^M f_M \,, \qquad \langle T^{ab} \rangle = 3 \rho^{-1} \omega_M{}^{ab} e^{4d/5} P_+{}^M{}_N f^N \,, \\
  \langle U_u \rangle &= \rho^{-1} e^{2d/5} b_u{}^M \xi_M \,, \qquad \langle P \rangle = \rho^{-1} e^{-8d/5} \theta \,, \\
  \langle R_{uvw} \rangle &= \rho^{-1} e^{2d/5} b_u{}^M b_v{}^N b_w{}^P f_{MNP} \,, \qquad \langle R_{uv}{}^{ab} \rangle = \rho^{-1} \omega_M{}^{ab} e^{4d/5} b_u{}^N b_v{}^P P_+{}^M{}_Q f_{NP}{}^Q \,,
 \end{split} \label{eq:TruncatedIntrinsicTorsion}
\end{equation}
where
\begin{equation}
 \begin{split}
  P_{-}{}^{MN} &= b_u{}^M b^{u,N} = \frac12 \left( \eta^{MN} - \gH^{MN} \right) \,, \quad \,\, P_+{}^{MN} = \eta^{MN} - b_u{}^M b^{u,N} = \frac12 \left( \eta^{MN} + \gH^{MN} \right) \,,
 \end{split}
\end{equation}
are left- and right-moving projectors. In the language of half-maximal gauged SUGRA one could say that the intrinsic torsion becomes the T-tensor, the ``flattened'' version of the embedding tensor.

We can now compute the scalar potential of the truncated theory. For this we will take the trombone tensor to vanish, i.e. $\xi_M = 0$. Otherwise, one does not obtain a consistent action principle and the gauged SUGRA would only be defined at the level of the equations of motion. We find
\begin{equation}
 \begin{split}
 \langle |e| V \rangle &= \rho^{5} |\hat{e}| \left[ - \frac14 e^{-2d} f_{MNP} f_{QRS} \left( \frac1{12} \gH^{MQ} \gH^{NR} \gH^{PS} - \frac14 \gH^{MQ} \eta^{NR} \eta^{PS} + \frac16 \eta^{MQ} \eta^{NR} \eta^{PS} \right) \right. \\
  & \left. \quad + \frac12 e^{-2d} f_M f_N \gH^{MN} + \frac18 e^{-6d} \theta^2 - \frac{\sqrt{2}}{12} e^{-4d} \theta \gH^{MNP} f_{MNP} \right] \,.
 \end{split}
\end{equation}
This is precisely the scalar potential of half-maximal gauged SUGRA, in particular with the singlet deformation parameter $\theta$ and the term $f^{MNP} f_{MNP}$ which vanishes by section condition but appears here automatically with the right relative coefficient.

We now see why it was crucial that the embedding tensor $f_{MNP}$, $f_M$, $\xi_M$ and $\theta_M$ are constant. Just as in the maximal case, see e.g. \cite{Hohm:2014qga}, this means that the $Y^{ab}$-dependence in the action factorises and thus any solutions of the lower-dimensional half-maximal gauged SUGRA can be uplifted to a solution of the full exceptional field theory, and thus, subject to solving the section condition, to 10- or 11-dimensional SUGRA. Another nice feature of the formulation given here is that if $\rho$, $\omega_M{}^{ab}$, $\hat{n}_a$ and $n^a$ satisfy the section condition, then the quadratic constraint of the half-maximal gauged SUGRA is automatically fulfilled. However, there are also be solutions of the quadratic constraint which violate the section condition, again analogously to the maximal case, for example \cite{Grana:2012rr,Lee:2015xga}.

\section{From exceptional field theory to heterotic double field theory}\label{s:Het}
The construction presented above can also be used to relate exceptional field theory to heterotic double field theory. This relation is reminiscent of the M-theory / heterotic duality \cite{Malek:2016vsh}.

\subsection{Ansatz and consistency condition}
We begin by considering exceptional field theory with an internal space admitting an $\SU{2}$-structure. We now expand this $\SU{2}$-structure in the same way as in our truncation Ansatz \eqref{eq:TruncationAnsatz}
\begin{equation}
 \begin{split}
  \langle A^a\rangle(x,Y) &= \rho^3(Y) n^a(Y) e^{-2d(x,Y)} \,, \qquad \langle\hat{A}_a\rangle(x,Y) = \rho^2(Y) \hat{n}_a(Y) \,, \\
  \langle B_u{}^{ab}\rangle(x,Y) &= \rho(Y) \omega_M{}^{ab}(Y) b_u{}^M(x,Y) \,, \qquad \langle\kappa\rangle(x,Y) = \rho(Y) e^{-2d(x,Y)/5} \,, \\
  \langle g_{\mu\nu}\rangle(x,Y) &= \rho^{2}(Y) \hat{g}_{\mu\nu}(x) e^{-4d(x)/5} \,,
 \label{eq:HetTruncationAnsatz}
 \end{split}
\end{equation}
but we now allow the ``scalar fields'' $b_u{}^M(x,Y)$ and $d(x,Y)$ to depend on both the external seven-dimensional coordinates $x^\mu$ \emph{as well as} the ten $Y^{ab}$, although we will soon restrict this dependence in a controlled manner. The fields $b_u{}^M(x,Y)$ become the left-moving frame fields of the heterotic double field theory and $d(x,Y)$ becomes the generalised dilaton of the theory. One proceeds similarly for the gauge fields $\TA_\mu{}^{ab}$, etc. We will call the theory thus obtain the ``gauged'' theory to differentiate it from the exceptional field theory we started off with.

The above Ansatz means that we do not obtain a truncated theory in seven dimensions, but still have a higher-dimensional theory. This procedure of making a truncation-like Ansatz but allowing the scalar fields to still depend on (some) of the $Y^{ab}$'s has recently been shown \cite{Ciceri:2016dmd} to reproduce the massive IIA theory \cite{Cassani:2016ncu} using exceptional field theory. We will show that the procedure here instead gives rise to the heterotic double field theory \cite{Siegel:1993th,Siegel:1993xq,Hohm:2011ex}, or heterotic supergravity once the section condition is solved.

We begin by imposing the same differential constraints on $\rho$, $\omega_M{}^{ab}$, $\hat{n}_a$ and $n^a$, i.e. \eqref{eq:DiffConstraints} as for consistent truncations. In order to make comparison with the heterotic double field theory of \cite{Hohm:2011ex} we will take
\begin{equation}
 f_M = \xi_M = \theta = 0 \,,
\end{equation}
although it is easy to consider a heterotic double field theory with some of these additional gaugings turned on. We will see that $f_{MNP}$ determines the gauge group of the heterotic double field theory. Note that this implies that the gauge group is necessarily a subgroup of $\ON{3,n}$.

We must also ensure that derivatives of the scalar fields $b_u{}^M(x,Y)$ and $d(x,Y)$ do not source doublets of $\SU{2}_S$. Additionally, just as in \eqref{eq:DiffConstraints}, we need to ensure that any excitations that they do source are captured in the truncation. We do this by imposing that
\begin{equation}
 \begin{split}
  \partial_{ab} d(x,Y) &= \frac12 \omega^M{}_{ab} \omega_M{}^{cd} \partial_{cd} d(x,Y) \,, \qquad \partial_{ab} b_u{}^M(x,Y) = \frac12 \omega^M{}_{ab} \omega_M{}^{cd} \partial_{cd} b_u{}^M(x,Y) \,, \label{eq:HetDiffConstraints}
 \end{split}
\end{equation}
and similarly for any other field of the gauged theory. In particular, this means that
\begin{equation}
 n^a \partial_{ab} d(x,Y) = n^a \partial_{ab} b_u{}^M(x,Y) = 0 \,,
\end{equation}
so that the theory effective has a six-dimensional internal space.

It is now natural to introduce the twisted derivatives
\begin{equation}
 D_M = \frac12 \tomega_M{}^{ab} \partial_{ab} \,,
\end{equation}
which we wish to identify with the $n+3$ derivatives of the heterotic double field theory. To do so, we require that they commute. However, the commutator is given by
\begin{equation}
 \begin{split}
  \left[ D_M, D_N \right] &= \frac{1}{2\rho} f_{MN}{}^P D_P - \frac32 \tomega_{[N}{}^{[cd} \partial_{cd} \tomega_{M]}{}^{ab]} \partial_{ab} \,.
 \end{split}
\end{equation}
This vanishes if impose the section condition as well as 
\begin{equation}
 f_{MN}{}^P D_P = 0 \,, \label{eq:fD=0}
\end{equation}
on any field in the gauged theory. This latter condition is also required from the perspective of the heterotic double field theory \cite{Hohm:2011ex}. Given that we take these conditions to be fulfilled we can now write the twisted derivatives as a partial derivative
\begin{equation}
 \partial_M = D_M \,.
\end{equation}

\subsection{Local symmetries of heterotic double field theory}
We can now compute the generalised Lie derivative acting on fields in the gauged theory. Consider two generalised vectors with the heterotic Ansatz \eqref{eq:HetTruncationAnsatz}
\begin{equation}
 \Lambda^{ab}(x,Y) = \rho(Y) \omega_M{}^{ab}(Y) \Lambda^M(x,Y) \,, \qquad V^{ab}(x,Y) = \rho(Y) \omega_M{}^{ab}(Y) V^M(x,Y) \,,
\end{equation}
with $\Lambda^M$ and $V^M$ satisfying the conditions \eqref{eq:HetDiffConstraints} and \eqref{eq:fD=0}. It is now straightforward to show that
\begin{equation}
 \gL_{\Lambda} V^{ab} = \rho \omega_M{}^{ab} \left( \Lambda^N \partial_N V^M - V^N \partial_N \Lambda^M + V^N \partial^M \Lambda_N + f_{NP}{}^M \Lambda^N V^P \right) \,, \label{eq:HetGL}
\end{equation}
which is the heterotic generalised Lie derivative with gauging determined by $f_{MNP}$.

Similarly the section condition
\begin{equation}
 \partial_{[ab} \Lambda^M \partial_{cd]} V^N = \frac{1}{4!} \epsilon_{abcde} n^e \eta^{PQ} \partial_P \Lambda^M \partial_Q V^N
\end{equation}
becomes the heterotic double field theory section condition. We thus recover the local symmetries and section condition of the heterotic double field theory.

\subsection{Intrinsic torsion and heterotic action}
Let us now evaluate the heterotic action and begin by calculating the intrinsic torsion. With \eqref{eq:HetTruncationAnsatz}, \eqref{eq:HetDiffConstraints} and \eqref{eq:DiffConstraints} one finds that the only non-vanishing components of the intrinsic torsion are
\begin{equation}
 \begin{split}
  U_u &= \rho^{-1} e^{2d/5} \Omega_u \,, \qquad R_{uvw} = \rho^{-1} e^{2d/5} \Omega_{uvw} \,, \\
  R_{uv}{}^{ab} &= \rho^{-1} \omega_M{}^{ab} e^{4d/5} e_{\bar{w}}{}^M \Omega_{uv}{}^{\bar{w}} \,, \qquad \kappa^{-2} \gL_{B_u} \left(\kappa U^u \right) = \rho^{-2} e^{4d/5} b_u{}^M \partial_M \Omega^u \,,
 \end{split}
\end{equation}
where we defined
\begin{equation}
 \begin{split}
  \Omega_u &= e^{2d} \partial_M \left( b_u{}^M e^{-2d} \right) \,, \qquad \Omega_{uvw} = L_{b_{[u}} b_v{}^M b_{w]M} \,, \qquad \Omega_{uv\bar{w}} = L_{b_{[u}} b_v{}^M b_{\bar{w}]M} \,,
 \end{split}
\end{equation}
exactly as in \cite{Hohm:2010xe}. Here $b_{\bar{u}}{}^M$ denote the $\bar{u} = 1, \ldots, n$ right-moving frame fields of the generalised metric, satisfying
\begin{equation}
 P_+{}^M{}_N b_{\bar{u}}{}^N = b_{\bar{u}}{}^M \,.
\end{equation}
Furthermore, we have used $L_\Lambda V^M$ to represent the heterotic generalised Lie derivative \eqref{eq:HetGL}
\begin{equation}
 L_{\Lambda} V^M = \Lambda^N \partial_N V^M - V^N \partial_N \Lambda^M + V^N \partial^N \Lambda_N + f_{MNP} \Lambda^N V^P \,. \label{eq:HetGL2}
\end{equation}

With these results one finds that the scalar kinetic terms \eqref{eq:ScalarKinTerms} become
\begin{equation}
 \langle |e| L_{kin} \rangle = \frac12 \rho^{5} e^{-2d} \hat{g}^{\mu\nu} \D_\mu \gH^{MN} \D_\nu \gH_{MN} + 4 \rho^{5} e^{-2d} \hat{g}^{\mu\nu} \D_\mu d \D_\nu d \,,
\end{equation}
where $\D_\mu = \partial_\mu - L_{A_\mu}$ represents the gauge-covariant derivative of the heterotic double field theory. Furthermore, the scalar potential reduces to
\begin{equation}
 \begin{split}
  \langle |e| V \rangle &= \rho^5 |\hat{e}| e^{-2d} \left[ b_u{}^M \partial_M \Omega^u + \frac12 \Omega_u \Omega^u - \frac1{12} \Omega_{uvw} \Omega^{uvw} - \frac14 \Omega_{uv\bar{w}} \Omega^{uv\bar{w}} \right. \\
  & \quad \left. - \frac14 \gH^{MN} \partial_M \hat{g}_{\mu\nu} \partial_N \hat{g}^{\mu\nu} \right] \,.
 \end{split}
\end{equation}
This matches the scalar kinetic terms and scalar potential of the heterotic double field theory in the frame formulation \cite{Hohm:2010xe,Hohm:2013nja}. One can similarly obtain the full action from exceptional field theory.

\section{Conclusions} \label{s:Conclusions}
Here we have shown how to use exceptional field theory to describe backgrounds admitting half the number of spinors. Such backgrounds admit a generalised $\SU{2}$-structure which can be defined by having certain non-vanishing generalised tensor fields, which can be thought of as the generalisation of complex and K\"ahler structure on K3 surfaces. We showed how to reformulate exceptional field theory in terms of the $\SU{2}$-structure thus making ${\cal N}=2$ SUSY manifest.

We then showed how one can define consistent truncations on such generalised $\SU{2}$-structure manifolds. The truncation Ansatz is made by expanding the $\SU{2}$-structure in terms of a set of modes which only depend on the internal space, with coefficients corresponding to seven-dimensional scalar fields. The consistency conditions are naturally encoded by the generalised Lie derivative and allow us to identify the embedding tensor of the half-maximal gauged SUGRA, including the singlet deformation parameter $\theta$. With the truncation Ansatz, the action reproduces the action of half-maximal gauged SUGRA.

Finally, we showed how one can use the methods described here to obtain the heterotic double field theory, using an Ansatz for all fields similar to the truncation Ansatz, but allowing the ``gauged'' fields to still depend on a subset of the 10 coordinates $Y^{ab}$. The intrinsic torsion of the $\SU{2}$-structure background now determines the gauge group of the heterotic theory. Furthermore, the generalised Lie derivative of exceptional field theory naturally gives rise to the heterotic double field theory Lie derivative, and the action reproduces that of the heterotic double field theory.

As we stressed in the introduction, dualities appear as ambiguities in exceptional field theory. The framework allows one to easily see when a lower-dimensional theory has two different, dual, higher-dimensional origins. With the results presented here one can also see when a lower-dimensional supergravity has a heterotic uplift in addition to type II. This happens for example, in K3 compactifications of 11-dimensional supergravity, where there is also an uplift to the heterotic SUGRA on $T^3$ with the gauge group broken to the Cartan subgroup \cite{Malek:2016vsh}.

The construction presented here can be generalised to other dimensions \cite{Malek:2017njj} as well as less SUSY, using the results of \cite{Ashmore:2015joa}. It would be interesting to apply these results to find new half-maximal consistent truncations which can amongst other things be used to find new vacua of 10- and 11-dimensional SUGRA. Another interesting aspect would be to study the moduli space of half-maximal vacua, similar to \cite{Ashmore:2016oug}.

\vskip1em

\noindent\textbf{Acknowledgements}: This article summarises a talk given at the ``Workshop on Geometry and Physics'', November 20th-25th 2016 at Ringberg Castle, Germany. The workshop was dedicated to the memory of Ioannis Bakas. The author would like to thank the organisers of the workshop for the opportunity to present this work. The author is supported by the ERC Advanced Grant ``Strings and Gravity" (Grant No. 320045).

\end{document}